\documentclass[twocolumn,pra,aps,showpacs]{revtex4}

\usepackage{mathptmx}
\usepackage{subfigure}
\usepackage{psfrag,graphicx}
\usepackage{epstopdf}
\usepackage{dcolumn}
\usepackage{amsmath,amssymb}
\usepackage{bm}
\usepackage{color}
\usepackage{latexsym}
\usepackage{color}
\usepackage[english]{babel}
\usepackage{latexsym}
\usepackage{subfigure}
\usepackage{amsmath}
\usepackage{amssymb}
\usepackage{amsfonts}
\usepackage{bm}
\usepackage{natbib}
\usepackage{appendix}
\usepackage{fancybox}

\def\ba{\begin{array}}
\def\ea{\end{array}}
\def\bi{\begin{itemize}}
\def\ei{\end{itemize}}

\def\be{\begin{equation}}
\def\ee{\end{equation}}
\def\bea{\begin{eqnarray}}
\def\eea{\end{eqnarray}}
\def\bse{\begin{subequations}}
\def\ese{\end{subequations}}
\def\bt{\begin{tabular}}
\def\et{\end{tabular}}
\def\bc{\begin{center}}
\def\ec{\end{center}}

\def\cphase#1#2{\left({#1}\right)_{#2}}
\def\cphaseBB#1#2{\text{B}^{(#1)}(#2)}
\def\cphaseBB#1#2{\text{B}_{#1}(#2)}
\def\cphasePB#1#2#3{\text{P}^{(#1)}_{#2}(#3)}
\def\cphasePB#1#2{\text{P}_{#1}(#2)}
\def\U{U}
\def\gateU#1#2{\U_{{#2}}\left({#1}\right)}
\def\gateUA#1#2{\U_{\text{A},{#2}}\left({#1}\right)}
\def\CPgateU#1#2{\U^{(#1)}(#2)}

\def\id{\mathbf{1}}
\def\halfpi{\tfrac{\pi}{2}}
\def\err{\epsilon}
\def\errA{\err_\text{A}}
\def\errA{\xi}
\def\ord{n}
\def\der{l}

\def\Ncp{N}

\def\F{F}
\def\target{\Theta}
\def\nqubits{d}
\def\ionphi{\zeta}

\def\R{\text{r}}
\def\B{\text{b}}
\def\ad{a^{\dagger}}
\def\phgate#1{\left[#1\right]}

\begin{document}

\author{Svetoslav S. Ivanov}
\affiliation{Department of Physics, St. Kliment Ohridski University of Sofia, 5 James Bourchier Blvd, 1164 Sofia, Bulgaria}
\author{Nikolay V. Vitanov}
\affiliation{Department of Physics, St. Kliment Ohridski University of Sofia, 5 James Bourchier Blvd, 1164 Sofia, Bulgaria}
\title{Composite two-qubit gates}

\begin{abstract}
We design composite controlled-phase gates, which compensate errors in the phase of a single gate.
The errors can be of various nature, such as relative, absolute or both.
We present composite sequences which are robust to relative errors up to the 6th order, with the number of the constituent gates growing just linearly with the desired accuracy,
 and we describe a method to achieve even higher accuracy.
We show that the absolute error can be canceled entirely with only two gates.
We describe an ion-trap implementation of our composite gates, in which simultaneous cancellation of the error in both the pulse area and the detuning is achieved.
\end{abstract}

\maketitle

\section{Introduction}\label{Sec-introduction}

Two-qubit gates form the basis of quantum computing: any quantum computation can be constructed entirely by these gates, combined with the one-qubit Hadamard and phase gates \cite{Nielsen2000}. Prominent examples are the controlled-phase (CPHASE) gate and the closely related controlled-NOT (CNOT) gate, which have been implemented on various physical platforms, such as ion traps \cite{traps}, nuclear magnetic resonance \cite{NMR}, photonic qubits \cite{photonic}, superconducting qubits \cite{supercond} and atoms \cite{atoms}.

Quantum computing, however, depends critically on the accuracy with which these gates are implemented and fault-tolerant computation is possible only if gate infidelity is very low, typically below $10^{-4}$ \cite{tolerance}.
Different sources of error can be identified in an experiment ranging from decoherence to imperfections in the control fields, most notably pulse length errors, field inhomogeneity, improper gate duration, frequency shifts, etc.
Because errors in the control fields lead to incorrect rotation angle of the state vector on the Bloch sphere, we refer to them as rotation errors.

Various techniques to deal with rotation errors have been proposed and implemented.
Composite pulses, for example, have found broad application, where systematic errors in the control field play a major role \cite{cpulses, BB1}.
A composite pulse is a sequence of pulses with well-defined phases, which are designed such that the errors from the constituent pulses largely cancel each other.
For example, the very popular broadband composite pulse BB1 of Wimperis \cite{BB1} cancels the error up to the second order on arbitrary state (without assumption of the initial state of the system).
More accurate pulses have been derived by Brown et al. \cite{Brown2004} but they have found limited application as they quickly become extremely long.
Low et al. \cite{Chuang2014} found optimal sequences yielding the same accuracy as Browns' but with much fewer primitive pulses.

In an important development, Jones \cite{Jones} showed how the available single-qubit composite pulses can be used to construct composite conditional two-qubit gates, which are the backbone of quantum computation.
In particular, he extended the BB1 pulse by Wimperis to construct a second-order broadband two-qubit gate, which is robust to rotation errors.
Later, Hill \cite{Hill} showed how to achieve robust CNOT gates from almost any interaction based on BB1 and used sequence concatenation for higher precision.
Gates of higher precision can be obtained even more efficiently by extending Low's sequences \cite{Chuang2014}.

In this work, we derive improved highly-accurate composite CPHASE gates of shorter duration and length than proposed so far.
Although derived for two qubits, these sequences can be used to construct robust multiqubit CPHASE gates as well.
In Sec.~\ref{sec:relative}, we present sequences, which cancel relative errors up to the 6th order using up to 12 gates, and describe how to obtain even higher accuracy.
In Sec.~\ref{sec:absolute}, we design sequences, which can handle errors of both relative and absolute nature.
Then, in Sec.~\ref{sec:ions}, we describe an implementation of our composite gates with linear laser-driven ion traps.
Remarkably, our gates can compensate simultaneous errors in the Rabi frequency and in the detuning.

\section{Compensation of relative error}\label{sec:relative}

\subsection{General framework}

An ideal two-qubit CPHASE gate, denoted by $\cphase{\theta}{}$, is represented by the propagator (in a rotated basis)
\be
\label{ideal}
\U(\theta) = e^{i \theta\sigma_x\sigma_x},
\ee
where $\sigma_x$ is the Pauli's $x$ matrix and $\theta$ is a rotation angle \cite{note}.
When two qubits are coupled with a coupling constant $J$ for a time $T$, then $\theta=J T$.
Relative rotation errors are described by multiplying $J$ or $T$ by an unknown factor, $1+\err$, so that $\theta$ is higher or lower than a desired value $\target$.
Therefore, rather than $\cphase{\target}{}$, in reality one obtains $\cphase{\target (1+\err)}{}$, represented by the propagator $\U(\target (1+\err))$.
The sensitivity of the gate to $\err$ can be much reduced by replacing single rotations with composed rotations,
\be
\label{def:seq1}
\phgate{\varphi_{\Ncp+1}}\cphase{\theta_\Ncp}{} \phgate{\varphi_\Ncp} \cdots \cphase{\theta_1}{} \phgate{\varphi_1} \cphase{\theta_0}{} \phgate{\varphi_0}.
\ee
Here time is running from right to left, so that the rightmost gate is the first one applied.
This sequence contains $\Ncp+1$ CPHASE gates $\cphase{\theta_k}{}$ and $\Ncp+2$ single-qubit phase gates $\phgate{\varphi_k}$, applied to a preselected qubit;
once chosen, the same qubit is used over the entire sequence. We will apply $\phgate{\varphi_k}$ on qubit 2.
The sequence \eqref{def:seq1} is represented by the following propagator
\be
\label{def:cgate1}
\CPgateU{\Ncp}{\target} = \F(\varphi_{\Ncp+1})\gateU{\theta_\Ncp}{} \F(\varphi_\Ncp) \cdots \gateU{\theta_0}{} \F(\varphi_0),
\ee
where
\be
\F(\varphi) = e^{-i \varphi \sigma_z}.
\ee

Note that by using the property \eqref{phasedgateprop} one can incorporate the phases in the $\U$-gates, yielding
\be
\label{def:cgate2}
\CPgateU{\Ncp}{\target} = \F(\phi_{\Ncp+1})\gateU{\theta_\Ncp}{\phi_\Ncp} \cdots \gateU{\theta_1}{\phi_1} \gateU{\theta_0}{\phi_0},
\ee
or $\CPgateU{\Ncp}{\target} = \F(\phi_{\Ncp+1}) \prod_{k=0}^\Ncp \gateU{\theta_k}{\phi_k}$, in a more compact form.
Here we have defined a phased CPHASE gate $\cphase{\theta}{\phi}$, represented by the propagator
\be
\label{phasedgate}
\gateU{\theta}{\phi} = e^{i \theta \sigma_x \sigma_\phi},
\ee
where $\sigma_\phi = \sigma_x \cos\phi + \sigma_y \sin\phi$, $\gateU{\theta}{0}=\gateU{\theta}{}$ and $\cphase{\theta}{0}=\cphase{\theta}{}$.
The realization \eqref{def:cgate2} may be more convenient in a practical setting, where the operator $\gateU{\theta}{\phi}$ is achieved at no additional cost
by a simple shift of the phase of the driving field, which does not represent a physical modification of the qubit.
We will use the realization \eqref{def:cgate2} for our composite sequences.
The following relations can be derived from the property \eqref{phasedgateprop}:
\be
\phi_l = -2\sum_{k=0}^{l-1}\varphi_k,\qquad
\phi_{\Ncp+1} = \sum_{k=0}^\Ncp \varphi_k,
\ee
with $l=1, 2, \ldots, \Ncp$.

Three families of composite sequences are generally considered: broadband, passband and narrowband, the first two of which will be of interest in the present work.

\subsection{Broadband sequences}

\subsubsection{General principles}

While every gate in the sequence \eqref{def:cgate2} is first-order sensitive to $\err$
(all angles $\theta_k$ are systematically wrong by some constant fraction $\err$),
the phases $\phi_k$ can be chosen such that the composite gate sequence is robust to $\err$ up to a certain higher order $\ord$.
To this end, we nullify the $\ord$ lowest-order propagator derivatives with respect to $\err$ by solving the following system of $n+1$ algebraic equations for the phases $\phi_k$:
\be\label{derivativesBB}
\left. \frac{\partial^\der}{\partial \err^\der} \left[\CPgateU{\Ncp}{\target}-\U(\target)\right]\right\vert_{\err=0} = 0,
\ee
with $\der = 0,1,\ldots, n$.
Such sequences exhibit robust profiles vs $\err$ around 0 and are called \emph{broadband}.
We denote them as $\cphaseBB{\ord}{\target}$ below.
Longer sequences provide more free parameters to vary ($\theta_k$ and $\phi_k$), thereby allowing to eliminate higher orders of $\err$.

Numerical calculations indicate that we must have $\theta_{0}=\target$ and $\phi_{0}=0$, and also that $\phi_{\Ncp+1}=0$ for all sequences with $\Ncp>2$.
Thus we are left with $2\Ncp-2$ parameters $\theta_k$ and $\phi_k$ to solve for.
It can be shown that for $k>0$ the angles $\theta_k$ can take values $\pi (s+1/2)$, where $s=0, 1, 2, \ldots$.
In what follows, we will restrict ourselves to $\theta_k=\halfpi$ or $\pi$, in order to minimize the total angle and thereby the total time duration of the composite sequence \eqref{def:cgate2}.
It follows from Eq.~\eqref{ExpToTrig} that $\gateU{\theta_k}{\phi_k}$ is equal to $\id$ for $\theta_k=\pi$ (even $s$) and to $i \sigma_x\sigma_{\phi_k}$ for $\theta_k=\halfpi$ (odd $s$). Thus it can be shown that the product $\prod_{k=1}^\Ncp \gateU{\theta_k}{\phi_k}$ applies $\sigma_x^m$ to qubit one, where $m$ counts the gates with $\theta_k=\halfpi$ in the product.
Because the zeroth-order approximation of $\cphaseBB{\ord}{\target}$ must reproduce $\gateU{\target}{}$, we must have $\prod_{k=1}^\Ncp \gateU{\theta_k}{\phi_k}=\id$ for $\err=0$.
This implies that $m$ is even, so that besides the target propagator $\gateU{\target}{}$, the sequence \eqref{def:cgate2} must contain an even number of $\halfpi$-gates.
Hence, the gate sequence \eqref{def:cgate2} acquires the form
\bse
\begin{align}
\label{def:cgate3a}
\phgate{\phi_{3}} \cphase{\halfpi}{\phi_{2}} \cphase{\halfpi}{\phi_{1}}\cphase{\target}{0}, \quad &\text{for $\Ncp=2$},\\
\label{def:cgate3b}
\cphase{\halfpi}{\phi_{\Ncp}} \cdots \cphase{\halfpi}{\phi_{2}} \cphase{\halfpi}{\phi_{1}}\cphase{\target}{0}, \quad &\text{for $\Ncp>2$},
\end{align}
\ese
having length of $\Ncp+1$ and a total angle of $\Ncp\halfpi+\target$.

We further found that in our sequences with $\Ncp>6$ and $\target=\tfrac{\pi}{4}$ we can set $\phi_1=0$.
The important implication from here is that we can merge the first two gates into a single gate $\cphase{\tfrac{3\pi}{4}}{}$. Note that, up to a global phase of $\pi$, this gate is equivalent to $\cphase{\tfrac{\pi}{4}}{\pi}$.
As a result, we can eliminate one gate $\cphase{\halfpi}{}$ from the sequences \eqref{def:cgate3b} simply by setting $\phi_0=\pi$, thereby yielding shorter sequences,
\be
\label{def:cgate4}
\cphase{\halfpi}{\phi_{\Ncp-1}} \cdots \cphase{\halfpi}{\phi_{2}} \cphase{\halfpi}{\phi_{1}}\cphase{\tfrac{\pi}{4}}{\pi}, \quad \text{for $\Ncp>6$},
\ee
with length $\Ncp$ and a total angle of $\Ncp\halfpi-\tfrac\pi4$.
We have subtracted 1 from each index for consistence of notation.

With the above assumptions the left-hand side of Eq. \eqref{derivativesBB} can be handled relatively easy by applying the identities \eqref{id}, \eqref{multinomial} and \eqref{derivatives2}, shown in Appendices \ref{Appendix:identities} and \ref{Appendix:derivatives}.
For $l=0$, we obtain
\be
\label{order0}
\sum_{k=0}^\Ncp (-1)^{k-\Ncp}\phi_k + \phi_{\Ncp+1} = 0.
\ee
Higher-order terms are not simple enough to be useful.

\subsubsection{Two-pulse sequence, $\ord=1$}

Let us consider a sequence with $\Ncp=2$.
Following the above arguments, we set $\theta_1=\theta_2=\halfpi$.
Zero- and first-order errors are cancelled by imposing the following set of equations:
\bse
\label{sys}
\begin{align}
-\phi_1+\phi_2+\phi_3 = 0, \\
\halfpi \left(e^{-\phi_1}+e^{-\phi_2}\right) +\target e^{i(\phi_1-\phi_2)}= 0.
\end{align}
\ese
We obtain $\phi_1=\phi$, $\phi_2=3\phi$ and $\phi_3=-2\phi$, where $\phi=\arccos(-\target/\pi)$, and the sequence is
\be
\cphaseBB{1}{\target} = \phgate{-2\phi}\cphase{\halfpi}{3\phi}\cphase{\halfpi}{\phi}\cphase{\target}{0}.
\ee
We consider this sequence to be of significant interest from experimental viewpoint for its reasonable robustness and small duration and length.

\subsubsection{Four-pulse sequence, $\ord=2$}

The composite gate with $\Ncp=4$ corrects for $\err$ up to the second order.
The following sequence is obtained
\be
\cphaseBB{2}{\target} = \cphase{\halfpi}{\phi}\cphase{\pi}{3\phi}\cphase{\halfpi}{\phi}\cphase{\target}{0}
\ee
with $\phi=\arccos(-\target/2\pi)$. The length is reduced to four gates as two adjacent phases are found to be equal.
This sequence coincides with the BB1 pulse derived by Wimperis \cite{BB1} and later used by Jones \cite{Jones} to construct a robust two-qubit gate, as discussed in the Introduction.

\subsubsection{Higher sequences, $\ord\geq3$}

\renewcommand{\arraystretch}{1.3}
\begin{table}[tb]
\centering
\begin{tabular}{l|c|l}
\hline
$\cphaseBB{\ord}{\target}$    & total angle   & phases $\phi_0,\phi_1,\phi_2,\phi_3,\ldots,\phi_\Ncp$ \\
\hline
$\cphaseBB{1}{\target}$       & $1.25\pi$              & 0, $\phi$, $3\phi$, $-2\phi$ [with $\phi=\arccos(-\theta/\pi)$]\\
$\cphaseBB{2}{\target}$       & $2.25\pi$              & 0, $\phi$, $3\phi$, $\phi$ [with $\phi=\arccos(-\theta/2\pi)$]\\
$\cphaseBB{3}{\pi/4}$         & $3.25\pi$              & 0, 1.725, 0.244, 1.127, 0.351, 1.785, 1.042\\
$\cphaseBB{4}{\pi/4}$         & $3.75\pi$              & 1, 0.170, 0.170, 1.374, 0.677, 1.598, 1.818,\\
                              &                        & 0.528, 1.995\\
$\cphaseBB{5}{\pi/4}$         & $4.75\pi$              & 1, 0.065, 2.257, 1.826, 1.020, 0.487, 1.452,\\
                              &                        & 1.671, 0.132, 0.812 \\
$\cphaseBB{6}{\pi/4}$        & $5.75\pi$               & 1, 2.193, 1.933, 0.737, 1.932, 1.286, 0.641,\\
                              &                & 1.531, 1.983, 1.240, 2.077, 0.579 \\
\hline
\end{tabular}
\caption{Broadband composite sequences $\cphaseBB{\ord}{\target}$,
 which cancel the relative error $\err$ up to order $\ord$, cf. Eqs. \eqref{derivativesBB}.
The phases for $\ord>2$ are given in units of $\pi$. The fidelities are shown in Fig. \ref{fig1}.
}
\label{tableBB}
\end{table}

Sequences of higher accuracy are calculated numerically (see Appendix \ref{Appendix:numerics}).
We picked a target angle of $\target=\pi/4$, which is traditionally used in quantum information to construct the CNOT gate \cite{Nielsen2000}.
Third, fourth, fifth and sixth orders in $\err$ are eliminated for $\Ncp=6$, 7, 9 and 11, respectively (cf. Eqs. \eqref{def:cgate3b} and \eqref{def:cgate4}), with corresponding total angles of $A=3.25\pi$, $3.75\pi$, $4.75\pi$ and $5.75\pi$.
The phases of the composite gates are given in Table \ref{tableBB}, where in places adjacent gates have identical phases.
Therefore, as for $\cphaseBB{2}{\target}$, we can combine these gates and reduce the overall length of the sequences.
To our knowledge, apart from $\cphaseBB{2}{\target}$, which coincides with BB1 by Wimperis \cite{BB1}, all broadband gates are original.

\begin{figure}[tb]
\includegraphics[width=0.95\columnwidth]{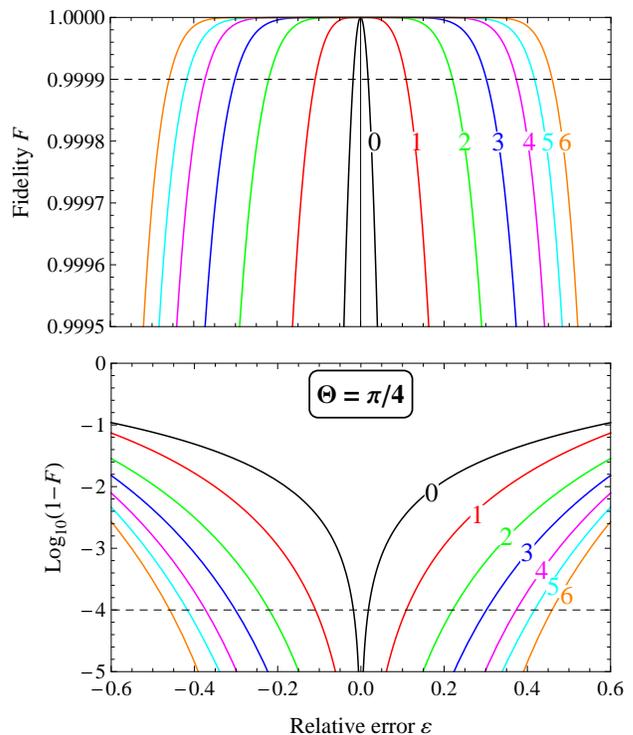}
\caption{
Fidelity of our broadband composite gates $\cphaseBB{\ord}{\target}$ to reproduce the gate $\exp \left(i \tfrac\pi4 \sigma_x \sigma_x\right)$, forming the basis of the CNOT gate, versus the relative error $\err$.
The order $\ord$ is displayed on each curve; 0 corresponds to a single uncorrected gate (cf. Eq. \eqref{ideal}).
The dashed line represents the 10$^{-4}$ benchmark level.
Note the dramatic increase of the ultrahigh fidelity range (infidelity below 10$^{-4}$) with $\Ncp$.
}
\label{fig1}
\end{figure}

The corresponding fidelities $F$ vs the error $\err$ are shown in Figure \ref{fig1},
where the standard definition of $F$ is used,
\be
F=\frac{\text{Tr}(A^\dagger B)}{\text{Tr} \left(A^\dagger A\right)}=\frac14 \text{Tr}(A^\dagger B)
\ee
with $A=U(\target)$ and $B=\cphaseBB{\ord}{\target}$.
Note that, as usual, a propagator infidelity of order $\err^{2\ord}$ corresponds to an error term in the underlying propagator of order $\err^{\ord}$.
A comparison with the benchmark of $1-10^{-4}$ (horizontal dashed line) reveals that a single gate can be fault-tolerant only if the error $\left|\err\right|$ does not exceed $1.8\%$.
The $\Ncp=2$ composite gate $\cphaseBB{1}{\pi/4}$ exhibits a clear improvement over the uncorrected single gate: the tolerance range is already $\left|\err\right|<11\%$.
As expected, the longer composite gates are fault-tolerant over a wider error range:
 we have $\left|\err\right|<22\%$ for $\cphaseBB{2}{\pi/4}$; $\left|\err\right|<30\%$ for $\cphaseBB{3}{\pi/4}$; $\left|\err\right|<37\%$ for $\cphaseBB{4}{\pi/4}$; $\left|\err\right|<42\%$ for $\cphaseBB{5}{\pi/4}$; $\left|\err\right|<46\%$ for $\cphaseBB{6}{\pi/4}$.
Our gates  $\cphaseBB{4}{\pi/4}$ and $\cphaseBB{6}{\pi/4}$ compare very well with the broadband gates BB4 and BB6 by Low et al. \cite{Chuang2014}: our $\cphaseBB{4}{\pi/4}$ with total angle $3.75\pi$ performs slightly better than Low's BB4 with total angle $4.25\pi$ and $\cphaseBB{6}{\pi/4}$ with total angle $5.75\pi$ performs slightly better than Low's BB6 with total angle $6.25\pi$.

The total angle, which determines the duration of the entire sequence, is $n \pi+\target$ (with $2n+1$ gates) for $n\leq 3$ and $(n-\tfrac12) \pi+\target$ (with $2n$ gates) for $3<n<7$, to correct the propagator to order $\ord$.
For example, $\cphaseBB{3}{\tfrac{\pi}{4}}$, which is third-order insensitive, has a total angle of $3 \pi+\target$.
Note for comparison that the same performance is achieved by the pulse B4,
 derived by Brown et al. \cite{Brown2004} and used by Jones \cite{Jones}, which requires 29 single gates and a total angle of $40\pi+\target$.

\subsection{Passband sequences}

Our broadband sequences are useful for the implementation of highly accurate gates on an isolated qubit pair.
In an actual experiment, however, it is possible that neighbouring qubits are involved in the interaction too, against our will, e.g. as a result of residual laser light addressing these qubits.
To suppress this effect while maintaining the robustness of the broadband sequences, one can use passband sequences.

Passband pulses satisfy the equations
\bse
\label{derivativesPB}
\begin{align}
\label{derivativesPBa}
&\left. \frac{\partial^{\der_1}}{\partial \err^{\der_1}} \left[\CPgateU{\Ncp}{\target}-\U(\target)\right]\right\vert_{\err=0} = 0,\\
\label{derivativesPBb}
&\left. \frac{\partial^{\der_2}}{\partial \err^{\der_2}} \left[\CPgateU{\Ncp}{\target}-\id\right]\right\vert_{\err=-1} = 0,
\end{align}
\ese
where ${\der_1}=0,1,\ldots, \ord_1$ and ${\der_2}=0,1,\ldots, \ord_2$.
Equations \eqref{derivativesPBa} define the broadband part around $\err=0$, and Eqs.~\eqref{derivativesPBb} define the narrowband part at $\err=-1$;
 the latter ensure that small coupling strengths, as ``felt'' by neighbouring qubits, yield negligible rotation.
These passband pulses, denoted as  $\cphasePB{\ord_1,\ord_2}{\target}$, are robust up to order $\ord_1$ around $\err=0$,
and up to order $\ord_2$ around $\err=-1$.
As such, passband sequences realize robust rotations upon our pair of qubits, as achieved using broadband sequences,
while suppressing rotations upon the remaining qubits \cite{laddressing}.

Eqs. \eqref{derivativesPB} can be handled relatively easy by applying the identities \eqref{id}, \eqref{multinomial} and \eqref{derivatives2}.
For Eqs. \eqref{derivativesPBa} we proceed as for the broadband sequences.
For $\der_2=0$, Eqs. \eqref{derivativesPBb} are automatically fulfilled, while for $\der_2=1$ and $\der_2=2$, they are reduced to
\bse
\begin{align}
2\target + \pi\sum_{k=1}^\Ncp e^{i\phi_k}=0,\\
3\pi^2-2\target^2+\pi^2\sum_{k<l=1}^\Ncp e^{i(\phi_k-\phi_l)}=0,
\end{align}
\ese
respectively, which are treated numerically, as it is the case also for $\der_2>2$.

\renewcommand{\arraystretch}{1.3}
\begin{table}[tb]
\centering
\begin{tabular}{l|l}
\hline\hline
$\cphasePB{\ord_1,\ord_2}{\target}$    & phases $\phi_0,\phi_1,\phi_2,\phi_3,\ldots,\phi_\Ncp$ \\
\hline
$\cphasePB{1,1}{\target}$       & 0, $\phi$, $-\phi$ [with $\phi=\arccos(-\target/2\pi)$]\\
$\cphasePB{2,1}{\target}$       & 0, $-\chi_1$, $-\chi_1+\chi_2$, $\chi_1+\chi_2$, $\chi_1-\chi_2$, $-\chi_1-\chi_2$, $\pi-\chi_1$\\
$\cphasePB{1,2}{\target}$       & 0, $\chi_1$, $\chi_1+\chi_2$, $-\chi_1+\chi_2$, $-\chi_1-\chi_2$, $\chi_1-\chi_2$, $\pi+\chi_1$\\
$\cphasePB{2,2}{\target}$       & 0, $\phi$, $-\phi$, $-\phi$, $\phi$ [with $\phi=\arccos(-\target/4\pi)$]\\
$\cphasePB{1,3}{\frac\pi4}$     & 0, 0.076, 1.604, 1.851, 0.595, 1.443, 0.751, 0.691, 1.111\\
$\cphasePB{3,3}{\frac\pi4}$     & 1, 0.091, 0.644, 1.866, 0.941, 1.596\\
\hline\hline
\end{tabular}
\caption{Passband sequences $\cphasePB{\ord_1,\ord_2}{\target}$,
which cancel the error $\err$ up to order $\ord_1$ around $\err=0$ and up to order $\ord_2$ around $\err=-1$ (cf. Eqs. \eqref{derivativesPB}).
The phases for $\cphasePB{1,3}{\target}$ and $\cphasePB{3,3}{\target}$ are given in units of $\pi$. The fidelities are shown in Fig. \ref{fig2}.
We have $\chi_1=\arccos\left(-\sqrt{\frac12+\frac{\target^2}{8\pi^2}}\right)$, $\chi_2=\arccos\left(-\sqrt{\frac{2\target^2}{4\pi^2+\target^2}}\right)$.
Note that $\cphasePB{2,1}{\target}$ is obtained from $\cphasePB{1,2}{\target}$ by a sign flip of $\chi_1$.
}
\label{tablePB}
\end{table}

\begin{figure}[tb]
\includegraphics[width=0.95\columnwidth]{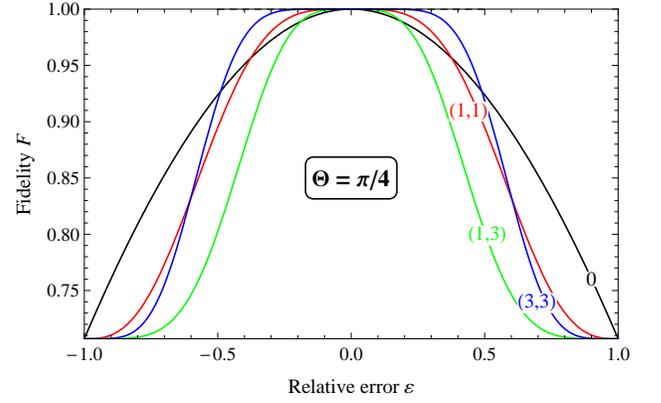}
\caption{
Fidelity of our passband composite gates $\cphasePB{\ord_1,\ord_2}{\tfrac\pi4}$
versus the relative error $\err$. The orders $(\ord_1,\ord_2)$ are displayed on each curve; 0 corresponds to a single uncorrected gate (cf. Eq. \eqref{ideal}).
}
\label{fig2}
\end{figure}

Calculated passband sequences correcting to different orders $\ord_1$ and $\ord_2$ are shown in Table. \ref{tablePB}.
The sequences are as follows: for $\cphasePB{1,2}{\target}$, $\cphasePB{2,1}{\target}$ and $\cphasePB{1,3}{\tfrac\pi4}$ (with total angles $3\pi+\target$, $3\pi+\target$ and $4.25\pi$) we use the sequence \eqref{def:cgate3b},
for $\cphasePB{1,1}{\target}$ and $\cphasePB{2,2}{\target}$ (with total angles $2\pi+\target$ and $4\pi+\target$) we have
\be
\cphase{\pi}{\phi_{\Ncp}} \cdots \cphase{\pi}{\phi_{2}} \cphase{\pi}{\phi_{1}}\cphase{\target}{0},
\ee
and for $\cphasePB{3,3}{\tfrac\pi4}$ (with total angle $5.75\pi$) we have
\be
\cphase{\pi}{\phi_{\Ncp}} \cdots \cphase{\pi}{\phi_{2}} \cphase{\pi}{\phi_{1}}\cphase{\tfrac{3\pi}{4}}{\pi}.
\ee
The corresponding fidelities for $\target=\tfrac\pi4$ are shown in Fig. \ref{fig2}.

Some of our sequences can be found in Ref. \cite{Chuang2014}: $\cphasePB{1,1}{\target}$ is identical to AP1, $\cphasePB{2,2}{\target}$ is identical to PD2 and PB1 by Wimperis \cite{BB1}. Note that $\cphasePB{3,3}{\tfrac\pi4}$ with total angle $5.75\pi$ performs almost as good as AP3 and PD4 with respective total angles $6.25\pi$ and $8.25\pi$.

\section{Compensation of absolute error}\label{sec:absolute}

In addition to the above sequences, which compensate relative errors in the target rotation angle $\target$, we have designed composite sequences, which suppress absolute errors that occur as a constant offset $\errA$ in the rotation angles, $\cphase{\target}{\phi}\rightarrow \cphase{\target + \errA}{\phi}$, represented by $\gateU{\target + \errA}{\phi}$. Like relative errors $\err$, these errors must enter systematically in the sequences.

Using the property $\gateU{\theta}{\pi}=\gateU{-\theta}{0}$, we have found that absolute errors can be eliminated \emph{completely} from $\gateU{\target}{\phi}$ with the sequence
\be
\cphase{\target}{\text{A},\phi} = \cphase{-\tfrac\target2}{\pi+\phi} \cphase{\tfrac\target2}{\phi},
\ee
represented by
\be
\gateUA{\target}{\phi} = \gateU{-\tfrac\target2}{\pi+\phi} \gateU{\tfrac\target2}{\phi}.
\ee
Indeed, we have
\begin{multline}
\gateUA{\target+\errA}{\phi} = \gateU{-\tfrac\target2+\errA}{\pi+\phi} \gateU{\tfrac\target2+\errA}{\phi}=\\
\gateU{\tfrac\target2-\errA}{\phi} \gateU{\tfrac\target2+\errA}{\phi} = \gateUA{\target}{\phi}.
\end{multline}
Potential errors in the phase $\phi$ can be removed following the composite technique of Ref.~\cite{Torosov}.

Finally, we construct composite gates robust to errors of either nature, relative and absolute.
This is done by substituting $\gateU{\theta}{\phi}$ with $\gateUA{\theta}{\phi}$ throughout in our sequences in Sec.~\ref{sec:relative}.
For example, a gate robust to $\err$ to the third order and to $\errA$ to any order is obtained with
\be
\phgate{-2\phi}\cphase{\halfpi}{\text{A},3\phi}\cphase{\halfpi}{\text{A},\phi}\cphase{\target}{\text{A},0},
\ee
where $\phi=\arccos(-\target/\pi)$.

Below we discuss a physical realization of our composite CPHASE gate with laser-driven linear ion traps.
While we consider ion traps, we note that our sequences are applicable to other systems, as well.

\section{Implementation with trapped ions}\label{sec:ions}

In trapped ions a popular two-qubit gate is the S{\o}rensen-M{\o}lmer (SM) gate \cite{Sorensen}.
It has been used by numerous ion trapping groups as a paradigmatic gate for quantum information processing.
The SM gate was demonstrated by Leibfried et al. \cite{Leibfried} and later by Kirchmair et al. \cite{Kirchmair} with fidelities around 97\%.
A record gate fidelity of 99.3\% has been achieved by Benhelm et al. \cite{Benhelm}.
Various dynamical decoupling techniques have been experimentally demonstrated to protect the two-qubit gate from the environment \cite{DD}.
Coherent error suppression using a pulse shaping technique has also been experimentally demonstrated \cite{Monroe}, where the effects of certain frequency and timing errors were suppressed.

In this section we will show how one can achieve a composite CPHASE gate, robust to i) rotation errors, which may be caused by improper laser intensity and timing, and ii) certain frequency errors, which may be caused by a shift in the trapping frequency.
The gate duration grows only linearly with the achieved precision contrary to previous proposals for coherent error suppression \cite{Monroe}, where exponential scaling is observed.

\subsection{Hamiltonian and propagator}

Consider two ions irradiated along the transverse $x$ direction with a bichromatic laser field with frequencies $\omega_{\R}=\omega_0-\omega_\text{cm}-\Delta$ and $\omega_{\B}=\omega_0+\omega_\text{cm}+\Delta$, tuned close to the first red ($\omega_{\R}$) and blue ($\omega_{\B}$) sidebands of a common vibrational mode.
Here $\omega_0$ is the frequency of the internal atomic transition of each ion, $\omega_\text{cm}$ is the frequency of the vibrational mode, and $\Delta$ is a suitably chosen detuning. The laser frequencies $\omega_\R$ and $\omega_\B$ sum up to twice the qubit transition frequency, while neither of the lasers is resonant to any level.
Thereby only transitions where the atomic states are changed collectively take place.
The interaction Hamiltonian is
\be
\label{Ham}
H = g\sum_{k=1}^2\sigma(\ionphi_k^+)
\left(\ad e^{i\Delta t - i\ionphi_k^{-}} + a e^{-i\Delta t + i\ionphi_k^{-}}\right),
\ee
where
$g$ is the (time-independent) Rabi frequency of the spin-phonon coupling and
$\sigma(\ionphi_k^+)=\sigma_k^{+}e^{-i\ionphi_k^{+}} + \sigma_k^{-}e^{i\ionphi_k^{+}}$, with $\sigma_k^+$ ($\sigma_k^-$) being the spin raising (lowering) operator for ion $k$.
The spin and the motional laser phases are defined by $\ionphi_k^{\pm} = \tfrac12(\ionphi_k^\B \pm \ionphi_k^\R)$, where $\ionphi_k^\B$ and $\ionphi_k^\R$ are, respectively, the laser phases of the blue- and red-detuned laser beams as seen by ion $k$.

The propagator $U$ is obtained using the Magnus expansion \cite{Magnus}:
\be
U = D(\alpha)\exp\left[i \frac{2g^2}{\Delta^2}\left(\Delta T - \sin\Delta T\right) \sigma(\ionphi_1^{+})\sigma(\ionphi_2^{+})\right],
\ee
where $T$ denotes the duration of interaction.
$D(\alpha)$ is a displacement operator, $D(\alpha) = \exp\left(\alpha \ad - \alpha^{\dagger} a\right)$ with
\be
\label{disp}
\alpha = -\frac{g T}{\Delta}\left(e^{i\Delta T}-1\right)\sum_{k=1}^2 \sigma(\ionphi_k^{+}) e^{-i\ionphi_k^{-}},
\ee
which causes an undesired change of the vibrational state of the ion system.

Now we discuss how to restore the vibrational state (eliminate $D(\alpha)$ from the propagator $U$),
while preserving the conditional dynamics, described by the $\sigma$-$\sigma$ term in $U$.
Note that if we only shift $\ionphi_k^{-}$ with $\pi$ in Eq. \eqref{disp},
we get a displacement of opposite magnitude, $D(\alpha)\rightarrow D(-\alpha)$.
This phase shift can be achieved either by a direct manipulation of the laser phase or by sandwiching $U$ with $\pi$ pulses on both ions.
Therefore, in order to restore the vibrational state, we apply a second bichromatic pulse of equal Rabi frequency $g$ and duration $T$ with a phase $\ionphi_k^{-}$ shifted with $\pi$.
Then the propagator becomes
\be
\label{finalpropagator}
U = \exp\left[i \frac{4g^2}{\Delta^2}\left(\Delta T - \sin\Delta T\right) \sigma(\ionphi_1^{+})\sigma(\ionphi_2^{+})\right],
\ee
provided that potential errors in the interaction variables are systematic.
Note that hereby we restore the vibrational state without even knowing the exact size of the detuning $\Delta$.

In the rest of the section we consider the implementation of both realizations \eqref{def:cgate1} and \eqref{def:cgate2}. The first allows us to
use global addressing, a key advantage in the S{\o}rensen-M{\o}lmer gate \cite{Sorensen}, at the expense of additional single-qubit phase gates, while the second requires individual addressing and possibly just a single phase gate (for $\Ncp=2$).

\subsection{Implementation with global addressing}

Global addressing implies equal laser phases for both ions, i.e. $\ionphi_k^{\pm}=\ionphi^{\pm}$.
Without loss of generality, we can assume that $\ionphi^{+}=0$, which implies that $\sigma(\ionphi_k^+) = \sigma_{x,k}$.
The propagator becomes
\be
U = \exp\left[i \frac{4g^2}{\Delta^2}\left(\Delta T - \sin\Delta T\right) \sigma_{x,1}\sigma_{x,2}\right],
\ee
which yields the gate $e^{i \theta\sigma_{x,1}\sigma_{x,2}}$ (cf. Eq. \eqref{ideal}) with $\theta$ given by
\be
\label{condition}
\theta = \frac{4g^2}{\Delta^2}\left(\Delta T - \sin\Delta T\right).
\ee

An important implication follows from here: potential systematic errors in the Rabi frequency $g$ (including unequal couplings), the detuning $\Delta$ and the pulse duration $T$
combine into a single error in the rotation angle, $\target\rightarrow \target(1+\err)$, which we already know how to suppress by using
our composite broadband sequences, listed in Table \ref{tableBB}.

When there are more than two ions in the trap, residual laser light is likely to couple neighbour ions, as well; neighbour ion $k$ will be coupled with Rabi frequency $g_k$.
As a result a rotation will occur with small angle $\theta_{k} = \frac{4 g g_k}{\Delta^2}\left(\Delta T - \sin\Delta T\right)$,
where we expect that $\theta_{k}\ll\target$.
This effect can be well suppressed by using our passband sequences, listed in Table \ref{tablePB}.

\subsection{Implementation with individual addressing}

Now the spin phase $\ionphi_2^+$ of ion two is modulated (relative to ion one), where the goal is to absorb the phase gates in the rotations [cf. sequences \eqref{def:cgate1} and \eqref{def:cgate2}].
Again, we set $\ionphi_k^{-}=\ionphi^{-}$ and without loss of generality, we assume that $\ionphi_1^{+}=0$.
The propagator \eqref{finalpropagator} becomes
\be
U = \exp\left[i \frac{4g^2}{\Delta^2}\left(\Delta T - \sin\Delta T\right) \sigma_{x,1}\sigma(\ionphi_2^+)\right],
\ee
which yields the gate $e^{i \theta\sigma_{x,1}\sigma(\ionphi_2^+)}$ [cf. Eq. \eqref{phasedgate}] that we need for the sequence \eqref{def:cgate2} with $\theta$ given by Eq. \eqref{condition}.

\section{Conclusion}

We have derived highly-accurate broadband and passband CPHASE gates, which correct rotation angle errors of relative and absolute nature.
For relative errors, the number of the ingredient gates and the duration of our sequences grow linearly
with the leading error order, as opposed to most proposals, where exponential growth is observed.
Absolute errors can be eliminated completely with a sequence of just two gates.
Implementation with trapped ions using bichromatic laser fields is discussed, where our sequences compensate errors both in the pulse area and the detuning.

\acknowledgments

This work is supported by the European Community’s Seventh Framework Programme (FP7/2007-2013) under Grant Agreement No. 270843 (iQIT).

\appendix

\section{Useful identities}\label{Appendix:identities}

From the identity $e^{-i {\phi} \sigma_z/2} \sigma_x e^{i {\phi} \sigma_z/2} = \sigma_\phi$,
we obtain
\be
\label{phasedgateprop}
\F_i(\phi/2)\gateU{\theta}{} \F_i(-\phi/2) = \gateU{\theta}{\phi},
\ee
where
$\F_i(\phi)=e^{-i \phi \sigma_{z,i}}$ and $i$ denotes a certain qubit.
For $i=2$ we have
\be
\label{ExpToTrig}
\gateU{\theta}{\phi} = e^{i\theta \sigma_x\sigma_\phi} = \cos\theta \id + i \sin\theta \sigma_x\sigma_\phi.
\ee

The following identities are useful for calculating the error terms
\bse
\label{id}
\be
\prod_{k=1}^{2l} \sigma(\phi_k) = \exp\left(i \sum_{k=1}^{2l}(-1)^{k}\phi_k \sigma_z \right),
\ee
\be
\prod_{k=1}^{2l+1} \sigma(\phi_k) = \sigma\left(-\sum_{k=1}^{2l+1} (-1)^k \phi_k\right),
\ee
\ese
where $l=0,1,2,\ldots$. For $l=1$ we have
\bse
\be
\sigma(\phi_1)\sigma(\phi_2) = \exp\left[-i(\phi_1-\phi_2)\sigma_z \right]
\ee
and
\be
\sigma(\phi_1)\sigma(\phi_2)\sigma(\phi_3) = \sigma(\phi_1-\phi_2+\phi_3).
\ee
\ese

\section{Calculation of propagator derivatives}\label{Appendix:derivatives}

To calculate the derivatives of the propagator in Eq. \eqref{derivativesBB} one can use the following property
\be
\label{multinomial}
\left. \frac{\partial^\der}{\partial \err^\der} \CPgateU{\Ncp}{\theta}\right\vert_{\err=0} = \sum_{\der_1+\ldots+\der_{\Ncp}=\der} \left(\begin{array}{c} \der \\ \der_1, \ldots, \der_\Ncp \end{array}\right)
\prod_{s=1}^{\Ncp}\left.\frac{\partial}{\partial \err^{\der_s}}\gateU{\theta_s}{\phi_s}\right|_{\err=0}.
\ee
Here the sum extends over all $\Ncp$-tuples ($\der_1,\ldots,\der_{\Ncp}$) of non-negative integers with $\sum_{s=1}^\Ncp \der_s = \der$.

For the derivatives we substitute
\be
\label{derivatives2}
\left.\frac{\partial^\der}{\partial \err^\der} \gateU{\theta(1+\err)}{\phi}\right|_{\err=0} = \theta^\der \gateU{\theta+\frac{\der\pi}{2}}{\phi}.
\ee

\section{Numerical procedure}\label{Appendix:numerics}
First, we construct a generic composite sequence of the form as shown in Eqs. \eqref{def:cgate1} or \eqref{def:cgate2}.
We calculate the derivatives from Eqs. \eqref{derivativesBB} or \eqref{derivativesPB} using the identities \eqref{multinomial} and \eqref{derivatives2}.
Then we proceed with a numerical minimization of the quantity
\be
D = \sum_{l=1}^\ord \left| \left. \frac{\partial^\der}{\partial \err^\der} \left[\CPgateU{\Ncp}{\target}-\U(\target)\right]\right\vert_{\err=0} \right|
\ee
where we use Newton's gradient-based method to determine the variables $\phi_k$ yielding $D=0$.
To minimize the number of the CPHASE gates $\gateU{\theta_k}{\phi_k}$, we start from a small number $\Ncp$, which is gradually increased, until we reach a solution to $D=0$.
Because we use a local optimization algorithm, we iteratively pick the initial values of the variables using a Monte-Carlo scheme.


\begin{thebibliography}{99}

\bibitem{Nielsen2000}
M. A. Nielsen and I. L. Chuang, \emph{Quantum Computation and Quantum Information} (Cambridge University Press, Cambridge, 2000).

\bibitem{traps}
F. Schmidt-Kaler, H. Haffner, M. Riebe, S. Gulde, G. P. T. Lancaster, T. Deuschle, C. Becher, C. F. Roos, J. Eschner, and R. Blatt, Nature (London) \textbf{422}, 408 (2003);
A. Khromova, Ch. Piltz, B. Scharfenberger, T. F. Gloger, M. Johanning, A. F. Var\'{o}n, and Ch. Wunderlich, Phys. Rev. Lett. \textbf{108}, 220502 (2012).

\bibitem{NMR}
L. M. K. Vandersypen and I. L. Chuang, Reviews of Modern Physics \textbf{76}, 1037 (2005).

\bibitem{photonic}
J. L. O'Brien, G. J. Pryde, A. G. White, T. C. Ralph and D. Branning, Nature \textbf{426}, 264-267 (2003);
S. Gasparoni, J.-W. Pan, P. Walther, T. Rudolph and A. Zeilinger, Phys. Rev. Lett. \textbf{93}, 020504 (2004);
P. Kok, W. J. Munro, K. Nemoto, T. C. Ralph, J. P. Dowling, and G. J. Milburn, Reviews of Modern Physics \textbf{79}, 135 (2007);
B. P. Lanyon, M. Barbieri, M. P. Almeida, T. Jennewein, T. C. Ralph, K. J. Resch, G. J. Pryde, J. L. O'Brien, A. Gilchrist and A. G. White, Nature Physics \textbf{5}, 134-140 (2009).

\bibitem{supercond}
T. Yamamoto, Yu. A. Pashkin, O. Astafiev, Y. Nakamura, J. S. Tsai, Nature \textbf{425}, 941-944 (2003);
J. H. Plantenberg, P. C. de Groot, C. J. P. M. Harmans and J. E. Mooij, Nature \textbf{447}, 836–839 (2007);
L. DiCarlo, J. M. Chow, J. M. Gambetta, Lev S. Bishop, B. R. Johnson, D. I. Schuster, J. Majer, A. Blais, L. Frunzio, S. M. Girvin and R. J. Schoelkopf, Nature \textbf{460}, 240-244 (2009).

\bibitem{atoms}
O. Mandel et al., Nature \textbf{425}, 937–940 (2003);
M. Anderlini et al., Nature \textbf{448}, 452–456 (2007).

\bibitem{tolerance}
P. Shor, Phys. Rev. A \textbf{52}, 2493 (1995);
A. M. Steane, Phys. Rev. Lett. \textbf{77}, 793 (1996).

\bibitem{cpulses}
M. H. Levitt, Prog. Nucl. Magn. Res. Spectr. \textbf{18}, 61 (1986);
H. H\"affner, C. F. Roos, and R. Blatt, Phys. Rep. \textbf{469}, 155 (2008);
N. Timoney, V. Elman, S. Glaser, C. Weiss, M. Johanning, W. Neuhauser, and C. Wunderlich, Phys. Rev. A \textbf{77}, 052334 (2008).

\bibitem{BB1}
S. Wimperis, J. Magn. Reson., Ser. A \textbf{109}, 221 (1994).

\bibitem{Brown2004}
K. R. Brown, A.W. Harrow, and I. L. Chuang, Phys. Rev. A \textbf{70}, 052318 (2004).

\bibitem{Chuang2014}
G. H. Low, T. J. Yoder, and I. L. Chuang, Phys. Rev. A \textbf{89}, 022341 (2014).

\bibitem{Jones}
J. A. Jones, Phys. Rev. A \textbf{67}, 012317 (2003);
J. A. Jones, Phys. Lett. A \textbf{316}, 24 (2003);
L. Xiao and J. A. Jones, Phys. Rev. A \textbf{73}, 032334 (2006);
W. G. Alway, J. A. Jones, J. Magn. Reson. \textbf{189}, 114–120 (2007).

\bibitem{Hill}
Charles D. Hill, Phys. Rev. Lett. \textbf{98}, 180501 (2007);
M. J. Testolin, C. D. Hill, C. J. Wellard, and L. C. L. Hollenberg, Phys. Rev. A \textbf{76}, 012302 (2007).

\bibitem{note}
The results of this work will remain unchanged if the $\sigma_y$ or the $\sigma_z$ basis is prefered, instead.
Also note that we can implement robust multiqubit CPHASE gates if the multiqubit CPHASE gate
$U^\nqubits\left(\theta\right) = e^{i \theta\sigma_x^{\otimes \nqubits}}$
is used as a base gate, instead of $\gateU{\theta}{}$ from Eq. \eqref{ideal}.

\bibitem{laddressing}
S. S. Ivanov and N. V. Vitanov, Optics Letters \textbf{36}, 1275-1277 (2011).

\bibitem{Torosov}
B. T. Torosov and N. V. Vitanov, Phys. Rev. A \textbf{90}, 012341 (2014).

\bibitem{Sorensen}
A. S{\o}rensen and K. M{\o}lmer, Phys. Rev. Lett. \textbf{82}, 1971 (1999);
A. S{\o}rensen and K. M{\o}lmer, Phys. Rev. A \textbf{62}, 022311 (2000).
K. M{\o}lmer and A. S{\o}rensen, Phys. Rev. Lett. \textbf{82}, 1835 (1999).

\bibitem{Leibfried}
D. Leibfried, B. DeMarco, V. Meyer, D. Lucas, M. Barrett, J. Britton,W. M. Itano, B. Jelenkovic, C. Langer, T. Rosenband, and D. J. Wineland, Nature (London) \textbf{422}, 412 (2003).

\bibitem{Kirchmair}
G. Kirchmair, J. Benhelm, F. Zahringer, R. Gerritsma, C. F. Roos, and R. Blatt, New J. Phys. \textbf{11}, 023002 (2009).

\bibitem{Benhelm}
J. Benhelm, G. Kirchmair, C. F. Roos, and R. Blatt, Nat. Phys. \textbf{4}, 463 (2008).

\bibitem{DD}
Ch. Piltz, B. Scharfenberger, A. Khromova, A. F. Var\'{o}n, and Ch. Wunderlich, Phys. Rev. Lett. \textbf{110}, 200501 (2013);
I. Cohen, S. Weidt, W. K. Hensinger and A. Retzker, New J. Phys. \textbf{17}, 043008 (2015).

\bibitem{Monroe}
D. Hayes, S. M. Clark, S. Debnath, D. Hucul, I.V. Inlek, K.W. Lee, Q. Quraishi, and C. Monroe, Phys. Rev. Lett. \textbf{109}, 020503 (2012).

\bibitem{Magnus}
W. Magnus, Commun. Pure Appl. Math. \textbf{7}, 649 (1954);
P. Pechukas and J. C. Light, J. Chem. Phys. \textbf{44}, 3897 (1966);
R. M. Wilcox, J. Math. Phys. \textbf{8}, 962 (1967);
S. Blanes, F. Casas, J. A. Oteo, and J. Ros, Eur. J. Phys. \textbf{31}, 907 (2010).

\end{thebibliography}
\end{document}